\documentstyle[preprint,aps,amsfonts]{revtex}
\begin{document}

\draft

\preprint{\vbox{\hbox{U. of Iowa preprint}}}

\title{Evidence for Complex Subleading Exponents
from the \\
High-Temperature Expansion of the Hierarchical Ising Model}

\author{Y. Meurice, G. Ordaz and V. G. J. Rodgers}

\address{Department of Physics and Astronomy\\
University of Iowa, Iowa City, Iowa 52246, USA}

\maketitle

\begin{abstract}
Using a renormalization group method, we calculate
800 high-temperature coefficients of the magnetic susceptibility
of the hierarchical Ising model.
The conventional quantities obtained from differences of ratios
of coefficients
show unexpected smooth oscillations with a period growing
logarithmically and can be fitted
assuming corrections to the scaling laws with complex exponents.

\end{abstract}

\newpage

The renormalization group method\cite{wilson}
has enhanced considerably our understanding
of elementary processes and critical phenomena.
In particular, it has allowed the computation of the critical
exponents of lattice models in various dimensions.
On the other hand, the critical exponents can be estimated from the analysis
of high-temperature series\cite{gaunt}.
Showing that the two methods give precisely the same answers
is a challenging
problem\cite{dis}. More generally, much could be gained if we
could combine these two approaches, in
particular, in the context of lattice gauge theories.

As far as the numerical values of the critical
exponents are concerned, there are two difficulties.
The first one\cite{nickel} is that
one needs {\it much} longer high-temperature series than
the ones available\cite{creutz}
(which do not go beyond order 25 in most of the
cases) in order to make precise estimates.
The second is that the
practical implementation of the renormalization group
usually requires projections into a manageable
subset of parameters characterizing the interactions.
It is nevertheless
possible to design lattice models\cite{dyson}, called hierarchical models,
which can be seen as approximate version of nearest neighbor models, for
which such projections are unnecessary.
For the hierarchical models,
the renormalization group transformation reduces to
a recursion formula which is
a simple integral equation involving  only the local measure.
This simplicity allows one to control rigorously\cite{sinai}
the renormalization group
transformation and to obtain accurate
estimates of the critical exponents\cite{epsi}.
As explained in the next paragraph, the recursion formula
also allows to calculate the high-temperature expansion to
very high order. Consequently, the hierarchical model
is well suited to study the questions addressed above.

In this letter, we report the results of
a large scale calculation which performs the high-temperature
expansion of the magnetic susceptibility of the hierarchical
Ising model up to order 800. The method of calculation
uses directly the renormalization group transformation
with a rescaling of the spin variable appropriate
to the study of the high-temperature fixed point.
This method has been presented in Ref.\cite{high}
and checked using results obtained with conventional
graphical methods\cite{jmp}.
For the sake of briefness, we shall follow exactly the set-up
and the notations of Refs.\cite{high,num} where the
basic facts concerning  the hierarchical model were reviewed.
The hierarchical
models considered here have $2^n$ sites, and the free
parameter $c$ which controls the strength
of the interactions is set equal to $2^{1-2/D}$ in order to approximate
a nearest neighbor model in $D$-dimensions.
We define the finite volume magnetic susceptibility
as
\begin{equation}\chi_n (\beta)=1\ + \ b_{1,n}\beta \ + \ b_{2,n}
\beta ^2 \ + \ ...
\end{equation}
In most of the calculations presented below, we have used $n$=100, which
corresponds to a number of sites larger than $10^{30}$.
{}From a mathematical point of view, the calculation of $\chi _n$
amounts to repeating $n$ times the Fourier transform
of the recursion formula\cite{high}
\begin{equation}
R_{n+1}(k)=C_{n+1}\exp(-{1\over 2}\beta
({c\over 2})^{n+1 }{{\partial ^2} \over
{\partial k ^2}})(R_{n}({k\over \sqrt{2}}))^2
\  ,
\end{equation}
expanded to the desired order in $\beta$, with the initial condition
$R_0=cos(k)$. The constant $C_{n+1}$ can be adjusted in such a way
that the Taylor expansion of $R_{n+1}(k)$ reads $1-(1/2)k^2 \chi _{n+1}+...$.
This calculation has been implemented with a C program and ran
for 6 weeks on a DEC-alpha 3000 in order to obtain 800 coefficients
for $D=3$.

For the discussion which follows, it is crucial to estimate
precisely the errors
made in the calculation of the coefficients. There are two sources
of errors: the numerical round-offs and the finite number of sites.
We claim that with $2^{100}$ sites and $3\leq D\leq 4$,
the finite volume effects
are several order of magnitude smaller than the round-off errors.
{}From Eq. (2), one sees that the leading volume dependence
will decay like $(c/2)^{n}$. This observation can be substantiated
by using exact results at finite volume\cite{jmp} for low order
coefficients, or by displaying the values of higher order
coefficients
at successive iterations as in Figure 1 of Ref.\cite{high}.
In both cases, we observed that the $(c/2)^{n}$ law worked
remarkably well.
For the main calculation presented below,
we have used $c=2^{1\over3}$ (i.e. $D=3$) and $n=100$, which gives
volume effects of the order of $10^{-20}$.
On the other hand, the round-off errors are expected to grow
like the square root of the number of arithmetical operations.
In Ref.\cite{high}, we have estimated that this number was approximately
$n.m^2$ for a calculation up to order $m$ in the high-temperature expansion
with $2^n$ sites. Putting this together, we estimated that for $n=100$,
the error on the $m$-th coefficient will be of order $m\times 10^{-16}$.
We have verified this approximate law by calculating the coefficients
using a rescaled temperature and undoing this rescaling after
the calculation.
We have chosen the rescaling factor to be 0.8482 and the rescaled
critical temperature is then approximately 1. This prevents the
appearance of small numbers in the calculation.
If all the calculations could be performed exactly,
we would obtain the same results as with the original method.
However, for  calculations with finite precision, the two calculations
have independent round-off errors.
Comparing the results obtained with the two methods for the
coefficients up to order 200 shows
that the numerical fluctuations of $b_m$ grow approximately
like $m \times 10^{-16}$. More conservatively, we can say that the numerical
errors
are bounded by $m\times 10^{-15}$.
We conclude that for the calculations reported below, the errors
on the coefficients are dominated by the numerical round-offs and
we estimate that they
do not exceed $10^{-12}$.

In order to estimate $\gamma $, we used standard methods described
in Refs.\cite{gaunt,nickel}. For the sake of definiteness, we
recall a few definitions.
First, we define $r_m=b_m/b_{m-1}$, the ratio of two
successive coefficients. We then define the normalized slope
$S_m$ and
the extrapolated slope $\widehat{S}_m $ as
\begin{eqnarray}
S_m & = & -m(m-1)(r_m - r_{m-1})/(mr_m -(m-1)r_{m-1})\ ;
\nonumber \\ & & \\
\widehat{S}_m & = & mS_m-(m-1)S_{m-1}\ .
\nonumber
\end{eqnarray}
The extrapolated slope, which is free of order $n^{-1}$
corrections\cite{nickel}, is displayed in Fig.1 for $m\leq 200$.
For comparison, we have also displayed the results for
$D=3.5$ and 4. A surprising feature is
the clear appearance
of large oscillations for $D=3$.
When $D$ is increased, the amplitude of these oscillations
diminishes. They are still present at $D=4$ and
can be seen better by plotting $\widehat{S}_{m+1}-\widehat{S}_{m}$.
One important point of this letter is to establish that these
oscillations are not due to the errors discussed above.
As a consequence of the multiplications by $m$ appearing in the definition
of the the extrapolated slope, the errors are amplified
by a factor which can be as large as $10^5$ for $m$ near 100
and $10^7$ for $m$ near 500. However, even when multiplied by
such a large factors our most conservative estimate of the
numerical errors gives errors on the
extrapolated slope which are several order of magnitude smaller
than the amplitude of the oscillations.
We have made independent checks of this statement for $D=3$ by
calculating directly the extrapolated slope for $n=$100 and 200
and by using an intermediate temperature rescaling
as explained above.
The smoothness of the oscillations appears clearly in Fig. 2,
where $\widehat{S}_m$ is displayed for $50\leq m\leq 800$. This
smoothness
rules out large numerical fluctuations.
In conclusion, we have established that the oscillations in the
extrapolated slope are a genuine feature of the model considered.
Fig. 2 also shows that the extrema are not equally spaced. Instead,
the location of one extremum can be approximately found by multiplying
the location of the previous extremum by 1.19. In other
words, the extrema of Fig. 2 would look equally spaced if the abscissa
variable
had been $ln(m)$ instead of $m$. This of course suggests the use
of a complex exponents since $Re(m^{i\sigma})=cos(\sigma ln(m))$.

In the conventional description\cite{parisi} of the renormalization group
flow near a fixed point with only one eigenvalue $\lambda_1 > 1$,
one expects that the magnetic susceptibility
can be expressed as
\begin{equation}
\chi=(\beta _c -\beta )^{-\gamma } (A_0 + A_1 (\beta _c -\beta)^{
\Delta } +....)\ ,
\end{equation}
with $\Delta=|ln(\lambda _2)|/ln(\lambda _1)$ and
$\lambda_2$ being the largest of the remaining eigenvalues.
It is usually assumed that these eigenvalues are real.
This implies\cite{nickel} that
\begin{equation}
\widehat{S}_m =\gamma -1 + B m^{-\Delta} + O(m^{-2}) \ .
\end{equation}
If $\Delta $ is real, there is no room for the oscillations in this
description.
Nevertheless,
the fact that the period of oscillation increases logarithmically
with $m$ suggests that one could modify slightly Eq. (5) by allowing
$B$ and $\Delta $ to be complex and selecting the real part of the modified
expression. This introduces two new parameters and we have chosen
to use the following modified parametrization of the extrapolated slope:
\begin{equation}
\widehat{S}_m =\gamma -1 + K m^{-\rho}cos(2\pi {ln(m/m_0) \over ln(\mu )})
+ O(m^{-2}) \ .
\end{equation}
This parametric expression allows good quality fits for $m$ large enough.
For instance, a least square fit for the $m\geq 300$ data,
yields $\mu = 1.412 $, $m_0 =512 $,
$\rho=0.67$, $\gamma $ = 1.310 and $K=2.53$.
The fit is displayed on Fig. 2.
More accurate results could presumably be obtained if we had a consistent
description of the oscillations involving definite relations among
the parameters of Eq. (6).
The value of $\gamma $ is in
good agreement with the result\cite{sinai,epsi} obtained with the
$\epsilon$-expansion, namely 1.300. The value of $\rho $ is not far
from
$|ln(\lambda _2)|/ln(\lambda _1)$ which is approximately 0.46 according
to Ref.\cite{sinai,epsi}.

We have considered two possible explanations of the oscillatory
behavior.
The first one is that one could replace $\lambda_2$ by a couple of
complex conjugated eigenvalues.
This possibility is not realized in any perturbative calculation we know.
For instance, the gaussian spectrum is real and its largest eigenvalues
are widely separated. One could imagine that when $D$
is continuously evolved from 4 to 3
for the hierarchical model, two real eigenvalues merge into each other
and subsequently evolve as complex conjugate of each other. However, the
fact that the oscillations persists at $D=4$, as indicated in Fig. 1,
goes against such
an explanation. Conformal theories in two dimensions provide examples
of calculations of the renormalization group eigenvalues as the
eigenvalues of the matrix of derivative of the beta functions.
For low order calculations, this matrix is symmetric\cite{zam} which implies
real eigenvalues.
A more attractive possibility, is that the susceptibility
of the hierarchical model satisfies a renormalization group
equation of the type discussed
in section II of Ref. \cite{nie}.
This equation allows corrections to the scaling law of the type
$(1+(A_1(\beta _c -\beta)^{i2\pi \over ln(\lambda )} +c.c)+..)$,
in the simplified case where only one eigenvalue $\lambda $ is considered.
Considering the values obtained with the fit we see that $\mu=1.412 $ is
close
to the the value of the largest eigenvalue\cite{sinai,epsi}
$\lambda_1$=1.427.
This result suggests that one should try
to find an
equation for the susceptibility of the hierarchical
model related to the one discussed in Ref.\cite{nie}.

In conclusion, we have shown that a calculational method of the
high-temperature expansion based on the renormalization group
method can be a very powerful tool when the hierarchical
approximation is used. It would be worth trying to improve this
method beyond this approximation. Our analysis of the magnetic
susceptibility has shown that unexpected oscillations appear
in the extrapolated slope. A detailed understanding
of these oscillations is required
in order to allow a precise comparison between the results obtained
from the high-temperature expansion and the $\epsilon$-expansion .

One of us (Y.M.), stayed at the Aspen Center for Physics during
the last stage of this work and benefited from stimulating
conversations with the participants, especially with N. Warner.

\newpage
\centerline{\bf Figure Captions}
\noindent
Fig. 1: The extrapolated slope $\widehat{S}_m $ for $m\leq 200$
and $D=3$, 3.5 and 4.

\noindent
Fig.2: The dots are the
extrapolated slope $\widehat{S}_m $ for $50\leq m\leq 800$
and $D=3$. The continuous curve is the fit described in the text.

\vfil
\end{document}